\documentclass[api,prl,showpacs,letterpaper,twocolumn,groupedaddress,footinbib,amssymb]{revtex4}
\usepackage{graphicx}
\usepackage[utf8x]{inputenc}
\usepackage{amsmath}
\usepackage{amssymb}
\usepackage{amsthm}
\usepackage{xcolor}

\begin{document}
\begin{abstract}
We calculate the real time non-equilibrium dynamics of quantum spin systems at finite temperatures.
The mathematical framework originates from the $C^*$-approach to quantum statistical mechanics and is applied to samples investigated
by means of spin-polarized scanning tunneling microscopy.
Quantum fluctuations around thermal equilibrium are analyzed and calculated.
The time averaged expectation values agree with the time averaged experimental data for magnetization curves.
The method is used to investigate the dynamics of a sample for shorter times than the resolution time of the experimental setup.
Furthermore, predictions for relaxation times of single spins on metallic and semiconductor surfaces are made.
To check the validity of our model we compare our results with experimental data obtained from Fe adatoms on InSb and Co adatoms on Pt(111) and find good agreement.
Approximated thermalization is found numerically for the expectation values of the spin operators.
\end{abstract}


\title{Non-equilibrium finite temperature dynamics of magnetic quantum systems: Applications to spin-polarized scanning tunneling microscopy }
\author{K.~Them}
\email[\textit{Electronic mail: }]
    {kthem@physnet.uni-hamburg.de}
\affiliation{Institute of Applied Physics and Microstructure
    Research Center, University of Hamburg,
    Jungiusstr.~11, 20355 Hamburg, Germany}
\author{T.~Stapelfeldt}
\affiliation{Institute of Applied Physics and Microstructure
    Research Center, University of Hamburg,
    Jungiusstr.~11, 20355 Hamburg, Germany}
\author{E.~Y.~Vedmedenko}
\affiliation{Institute of Applied Physics and Microstructure
    Research Center, University of Hamburg,
    Jungiusstr.~11, 20355 Hamburg, Germany}
\author{R.~Wiesendanger}
\affiliation{Institute of Applied Physics and Microstructure
    Research Center, University of Hamburg,
    Jungiusstr.~11, 20355 Hamburg, Germany}
\date{\today}
\maketitle

\section{Introduction}
Spin sensitive studies of individual magnetic adatoms and atomic ensembles on surfaces with spin-polarized scanning tunneling microscopy
(SP-STM) \cite{WiesendangerAnfang,Wiesendanger:RMP2009} have raised the necessity of a quantum-mechanical description of the spin dynamics evoked by SP-STM experiments.
Magnetization curves obtained in experiments are typically described using the expectation values of observables using a time independent, i.e. kinematic,
Gibbs ensemble average \cite{Meier:Science(2008),Kha:Nature2010}. However, an SP-STM measurement is a time-average of the orientation of a spin component selected
by the given spin orientation of the probe tip.
Therefore, the dynamics of the magnetization in the sample under the study remains unknown within the experimental resolution time.
It would be helpful to compensate this lack of knowledge
with theoretical investigations.

When the STM tip comes towards an atom or a cluster under the study, the Hamilton operator
of the system changes due to the interactions with
tunneling electrons. The perturbed dynamics drives the state out of equilibrium and the ergodicity is not a priory ensured.
Therefore, the ergodicity of a system has to be checked for a reliable interpretation of experimental results.
A still unexplained finding is the extremely high switching frequency of Co atoms on Pt(111) at zero magnetic field \cite{Meier:Science(2008)}.
In contrast to magnetic atoms on insulating substrates Co/Pt(111) possesses very strong out-of-plane anisotropy (9 meV) without any transversal contributions.
Hence, the Hamiltonian of the free system is diagonal in the $|S_z\rangle$ basis and, therefore, the tunneling rate is zero \cite{Meier:Science(2008)}.
The anisotropy barrier is approximately 100 times larger than the temperature used in experiments. Therefore, the Boltzmann probability to pass this barrier by
thermal activation is also negligible.
The measured magnetization, in contrast, is zero at zero field even in the regime of elastic tunneling, where the tunneling current
density is minimal.
A related problem is the so-called "return to equilibrium", also referred as relaxation. The relaxation of an excited system depends
on its initial state \cite{Bratelli}.
There might exist some initial states from which the system can return to equilibrium and there
might exist some other initial states from which this process will not happen.

The formal, mathematical description of systems which can be described by different Hamilton operators before/after and during the measurement process has been
elaborated in the framework described in \cite{Bratelli}.
The algebraic approach to quantum statistical mechanics provides appropriate mathematical tools to
verify the dynamical relaxation of a disturbed system (return to equilibrium) analytically \cite{Bratelli,ArakiI,ArakiII}.
Several theoretical aspects in the algebraic approach to quantum spin systems, such as propagation velocities, are of actual
research interest \cite{NachtergaeleI}-\cite{NachtergaeleIIIIIII}.
Particularly, it is proven that a
mathematically exact return to equilibrium is ensured if the dynamical system satisfies some form of
asymptotic abelianness and if the initial state is a perturbed KMS state (Kubo-Martin-Schwinger) \cite{Bratelli}.
In the present paper we use the algebraic formulation of quantum statistical mechanics \cite{Bratelli} to clearly separate the
thermal equilibrium Gibbs states and the time evolution of the system during SP-STM experiments.

It is generally believed, that only large systems show a relaxation process.
We demonstrate that also expectation values of relatively small quantum spin systems, containing less than 10 particles, return approximately to equilibrium
when a perturbed KMS state is used as an initial state.
Especially interesting is the theoretical analysis of the dynamics on time scales which are not accessible for an STM.
Using exact diagonalization we
calculate the dynamics of single quantum spins during and after SP-STM measurements at finite temperatures. We demonstrate that the relaxation times of those quantum
objects on different substrates lie in the femto-, pico- or nanosecond regime.
It means that a good approximation of the long time behavior can be given for certain classes of real finite systems.
To check whether the short time dynamics has a reliable behavior, the calculated relaxation time has been compared with experimentally
determined life times for single spins \cite{Kha:Nature2010}.
Ground states are obtained from KMS states if the zero temperature limit is performed.

\section{Method}
The SP-STM set-up is approximated by (in general) two different Hamiltonians in our approach. There is a Hamiltonian $H$ for the free system
and, if a measurement is started, we get an additional hermitian operator $P$ for the interaction between the tip and the sample.
Hence, if the tip is moved towards the surface, the system switches from $H$ to $H+P$
because of the sudden emergence of tunneling electrons causing the interaction between the tip and the sample.
\begin{figure}[h]
 \includegraphics*[scale=0.8,viewport=102 641 375 791]{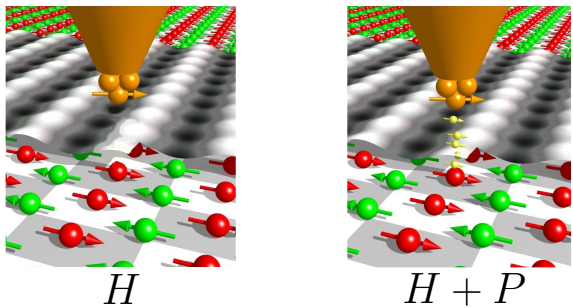}
 \caption{$H$ describes the free system and $H+P$ describes the perturbed system, i.e. the system with interaction between the tip and the sample.}
\end{figure}
The STM-tip can be used to prepare a system with desired expectation values.
For the corresponding state we choose a perturbed KMS state $\omega^{\beta P}$.
The reason for the application of this state is justified in the following text.

To describe a quantum spin system we consider particles on a lattice $\mathbb{Z}^d$ and associate with each point $x\in\mathbb{Z}^d$ a
Hilbert space $\mathcal{H}_{x}$ of dimension $2s(x)+1$. With a finite subset $\lambda\subset\mathbb{Z}^d$ we associate the
tensor product space $\mathcal{H}_\lambda=\bigotimes_{x_i\in\lambda}\mathcal{H}_{x_i}$.
The local physical observables are contained in the algebra of all bounded operators acting on $\mathcal{H}_\lambda$. This
is the $C^*$-algebra $\mathcal{A}_\lambda\cong\bigotimes_{x_i\in\lambda}M_{2s(x_i)+1}$
in which $M_n$ denote the algebra of $n\times n$ complex matrices. Physically, this can be interpreted as follows: at each lattice site $x$
there is a particle with spin quantum
number $s(x)$ and with $n=2s(x)+1=dim(\mathcal{H}_{x})$ degrees of freedom.
The numerical calculations are done for systems of finite dimensions. The indices $x$, $x_i$ and $\lambda$ are suppressed
in the following text for clarity.
A mixed (or normal) state $\omega$ is described as a normalized positive linear functional over the matrix algebra $\mathcal{A}$
and is given by a density matrix $\rho$.
\begin{equation}
 \omega :\mathcal{A}\rightarrow \mathbb{C},  \quad A\mapsto \omega(A)=Tr(\rho A)
\end{equation}
The dynamical evolution of an observable $A\in\mathcal{A}$ for a system with Hamiltonian $H=H^*\in M_n$ can be described by the Heisenberg relations
\begin{equation}
 \tau_t:\mathcal{A}\rightarrow \mathcal{A} , \quad A\mapsto \tau_t(A)=e^{\frac{itH}{\hbar}}A\,e^{-\frac{itH}{\hbar}}.
\end{equation} Thus, the map $t\in\mathbb{R}\mapsto\tau_t$ is a one-parameter group of $^*$-automorphisms of the matrix algebra $\mathcal{A}$.
In our formalism the Hamiltonian $H$ describes a "free" quantum system without any interaction with the magnetic tip (Fig. 1, left).
When the spin-polarized current starts to flow through the system under investigation, the interaction between the tip and the sample is described by
the perturbed Hamiltonian $H+P$ (Fig. 1, right).
A perturbed dynamical evolution can be introduced by
\begin{equation}
 \tau_t^P:\mathcal{A}\rightarrow \mathcal{A} , \quad A\mapsto \tau_t^P(A)=e^{\frac{it(H+P)}{\hbar}}A\,e^{-\frac{it(H+P)}{\hbar}}.
\end{equation}
Thermal equilibrium at inverse temperature $\beta$ is modeled by the Gibbs canonical ensemble state which is also the unique
$(\tau,\beta)$-KMS state, denoted by $\omega^\beta$ and given by
\begin{equation}
 \omega^\beta(A)=\frac{Tr(e^{-\beta H}A)}{Tr(e^{-\beta H})}.
\end{equation} These states are invariant under the action of $\tau$, i.e. $\omega^\beta(\tau_t(A))=\omega^\beta(A)$, but in general not
invariant under the action of $\tau^P$. A corresponding perturbed $(\tau^P,\beta)$-KMS state can be introduced by
\begin{equation}\label{eq:calcprob}
 \omega^{\beta P}(A)=\frac{Tr(e^{-\beta (H+P)}A)}{Tr(e^{-\beta (H+P)})}.
\end{equation}
Now we can plug the perturbed dynamics into the unperturbed equilibrium state \cite{Bratelli}
\begin{equation}\label{eq:measure}
 \omega^\beta(\tau_t^P(A))\equiv\langle A\rangle_1(t).
\end{equation} The brackets $\langle ...\rangle(t)$ shall mean that we calculate the time evolution of an expectation value for the observable $A$.
This corresponds to the situation when the spin-polarized tunneling current is switched on at the time $t=0$ and the system was in thermal equilibrium for $t<0$.
The function (\ref{eq:measure}) is used to model the process of a measurement of a magnetization curve.
We can also plug the unperturbed dynamics into the perturbed equilibrium state \cite{Bratelli}
\begin{equation}\label{eq:r}
\omega^{\beta P}(\tau_t(A))\equiv\langle A\rangle_2(t).
\end{equation}
In this case a spin-polarized current is switched off at the time $t=0$.
The state $\omega^{\beta P}$ can be prepared with SP-STM.
The function (\ref{eq:r}) can also be used to model the process of return to equilibrium.
If a certain model Hamiltonian is associated with $H$ and $P$, the evaluation of expectation values with (\ref{eq:measure}) and (\ref{eq:r})
can be calculated with different numerical methods. Some other examples to which this approach can be applied can be found in
\cite{Potthoff}-\cite{Stapelfeldt}.
To make connection to the more common theoretical models for SP-STM, we notice that $P$ could, for example, be given by a kind of s-d interaction or
the Tersoff-Hamann model. The choice of $P$ to be a local magnetic field is appropriate to save memory, which is needed by the calculation of a relaxation process.

A measurement in SP-STM is a time average over a time period $\Delta t$. For example the time resolution of the measurement in \cite{Meier:Science(2008)} is
$\Delta t=10$ ms. Each point on a measured magnetization curve corresponds then to the value:
\begin{equation}\label{eq:measurement}
 \langle A\rangle_{\Delta t}=\frac{1}{\Delta t}\int^{\Delta t}_0dt\, \omega^\beta(\tau^P_t(A)),
\end{equation} where $A$ is a spin component, i.e., $S_x,S_y$ or $S_z$.

It is worth to mention, that for infinite systems the equations (\ref{eq:measure}) and (\ref{eq:r}) are widely analysed in mathematical physics \cite{Bratelli},
but it seems that they were never applied to real physical spin systems. If the substitution $H\rightleftarrows H+P$ is replaced by
$(\mathcal{A},\tau)\rightleftarrows(\mathcal{A},\tau^P)$, comprehensive mathematical structures \cite{Bratelli}-\cite{NachtergaeleIIIIIII},
\cite{RobinsonI}-\cite{Y. Ogata} can be applied for the
analysis of physical systems.
If any form of asymptotic abeliannes is satisfied one finds
\begin{equation}
 \lim_{t\rightarrow\infty}\omega^\beta(\tau_t^P(A))=\omega^{\beta P}(A),
\end{equation} which motivates the application of perturbed KMS states as states which can be prepared with an SP-STM.
On the other hand
\begin{equation}
 \lim_{t\rightarrow\infty}\omega^{\beta P}(\tau_t(A))=\omega^{\beta }(A),
\end{equation} which motivates the application of (\ref{eq:r}) to calculate a relaxation process after the spin current has been switched off.
The states $\omega^\beta$ and $\omega^{\beta P}$ are related by the \textit{M\o{}ller morphisms} $\gamma_{\pm}$, especially if
$L^1(\mathcal{A}_0)$-asymptotic abelianness
is satisfied \cite{Bratelli}. Furthermore, one finds that ground states, which are zero temperature KMS states, have a tendency to be
less stable than KMS states at finite temperature.

\section{Numerical calculations}
Magnetic adatoms on a metallic or a semiconductor surface can often be modeled with a Hamiltonian of the form
\begin{equation}\label{eq:HamiltonianRelax}
 H=\sum_i\big(DS^2_{iz}+E(S^2_{ix}-S^2_{iy})\big)+\sum^n_{i,j,\alpha}J^\alpha_{ij}S^\alpha_j\sigma^\alpha_i.
\end{equation} The first term of the Hamiltonian describes the magnetic properties of adatoms. The second summation approximates the interaction of magnetic atoms
with substrate electrons and is sometimes called s-d interaction.
$S^\alpha$ are the components $\alpha=x,y,z$ of the spin operators of the adatoms and $\sigma^\alpha_i$ are Pauli matrices
corresponding to the spin components of the substrate electrons at a
lattice site $i$. $J^\alpha_{ij}$ is the strength of the Heisenberg interaction between the adatom and the substrate electrons.
The strength of $|J^\alpha_{ij}|$ in Eq. (\ref{eq:HamiltonianRelax}) has been distributed randomly between $0$ and $0.8$ meV,
corresponding to the typical strength of exchange interaction
between magnetic adatoms on conducting or semiconducting surfaces.
For our model calculations two different types of perturbation were chosen:
\begin{equation}
 P=\sum_{i,\alpha}g\mu_BB^\alpha S^\alpha_i,
\end{equation} where $B^\alpha$ is a local magnetic mean field acting on the sample, $g$ a gyromagnetic constant and $\mu_B$ is the Bohr magneton. Alternatively one can choose
\begin{equation}
 P=\sum_{i,j,\alpha}J^{'\alpha}_{ij} S^\alpha_i\sigma^\alpha_i+\sum_{i,\alpha}m^\alpha\sigma^\alpha_i,
\end{equation} where $\sigma^\alpha_i$ are the spin operators of the tunneling electrons and $m^\alpha$ is the magnetization of the tip.
The values of $J^{'\alpha}_{ij}$ have been chosen to be similar to those of $J^\alpha_{ij}$.

In the first part of calculations we study a single adatom coupled to bath electrons. It is a priori not clear, whether the described
finite quantum system is able to approach its equilibrium.
It will be demonstrated that already $n=8$ substrate (or bath) electrons acting as a heat bath are sufficient, for a single adatom at zero magnetic field, to reach
thermal equilibrium, when a perturbed KMS state is used.
After a characteristic time $t_0$ the expectation value $\langle S^\alpha\rangle_2(t)$ relaxes and fluctuates around its thermal equilibrium value, i.e.,
\begin{equation}
 \omega^{\beta P}(\tau_t(S_z))\longrightarrow\approx\omega^\beta(S_z)=\frac{Tr\big(e^{-\beta H}S_z\big)}{Tr\big(e^{-\beta H}\big)}.
\end{equation} The amplitude and the form of fluctuations are temperature dependent. To make sure that we get realistic relaxation times, we compare our calculations
with the life-time of an Fe adatom on InSb estimated in recent SP-STM experiments \cite{Kha:Nature2010,Kha:Iterniant}.
The corresponding parameters $g_{Fe}=2$, $D=-1.4$ meV, $E=0.22$ meV and $S=1$ for the iron atom are taken from \cite{Kha:Nature2010}.
To calculate the relaxation time we use the expression $\omega^{\beta P}(\tau_t(S_y))$, in which the time evolution is
generated by the Hamiltonian Eq. (\ref{eq:HamiltonianRelax}). The $y$-component of the iron atom was investigated in this experiment \cite{Kha:Nature2010}.
In Fig. 2 the agreement of the interpretation of $\omega^{\beta P}(\tau_t(S_y))$ as a relaxation process with experimental data \cite{Kha:Nature2010,Kha:Iterniant} is verified.

As one can see from Fig. 2, the expectation value of the magnetization increases from -1 to zero and then fluctuates around thermal equilibrium.
The experimental estimation of the lifetime $t_{l.t.}$ of the excited state was done by the formula $t_{l.t.}=\frac{\hbar}{2\Delta E}$, where $\Delta E$ is the energy
difference between the states, obtained from inelastic SP-STM\cite{Kha:Nature2010,Kha:Iterniant}.

\begin{figure}[h]
\includegraphics*[scale=0.68,viewport=149 553 510 670]{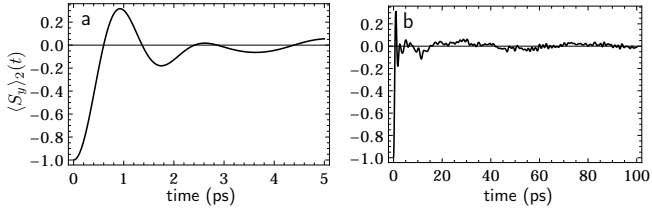}
\caption{Return to equilibrium for the spin component $S_y$ of an iron atom on an indium antimonide surface for short and long times at $T=4.2$ K.
Fig. 2 a): The experimentally estimated life time of $800$ fs is in good agreement with the
calculated relaxation process. Fig. 2 b): After the relaxation is done the expectation value remains near thermal equilibrium $\omega^\beta(S_y)=0$.}
\end{figure}

In Fig. 3) and 4) the calculated functions for a high temperature of $100$ K and a low temperature of
$4$ K are shown. The short time and the long time behavior are analyzed for different values of the parameters $E$ and $D$ in Eq. (\ref{eq:HamiltonianRelax}).
In Fig. 3 a)-d) the function $\omega^{\beta P}(\tau_t(S_z))$ is plotted for different values of $E$ and fixed $D$, while $E$ is fixed and $D$ is varied in Fig. 4 a)-d).
\begin{figure}[h]
 \includegraphics*[scale=0.68,viewport=149 441 511 672]{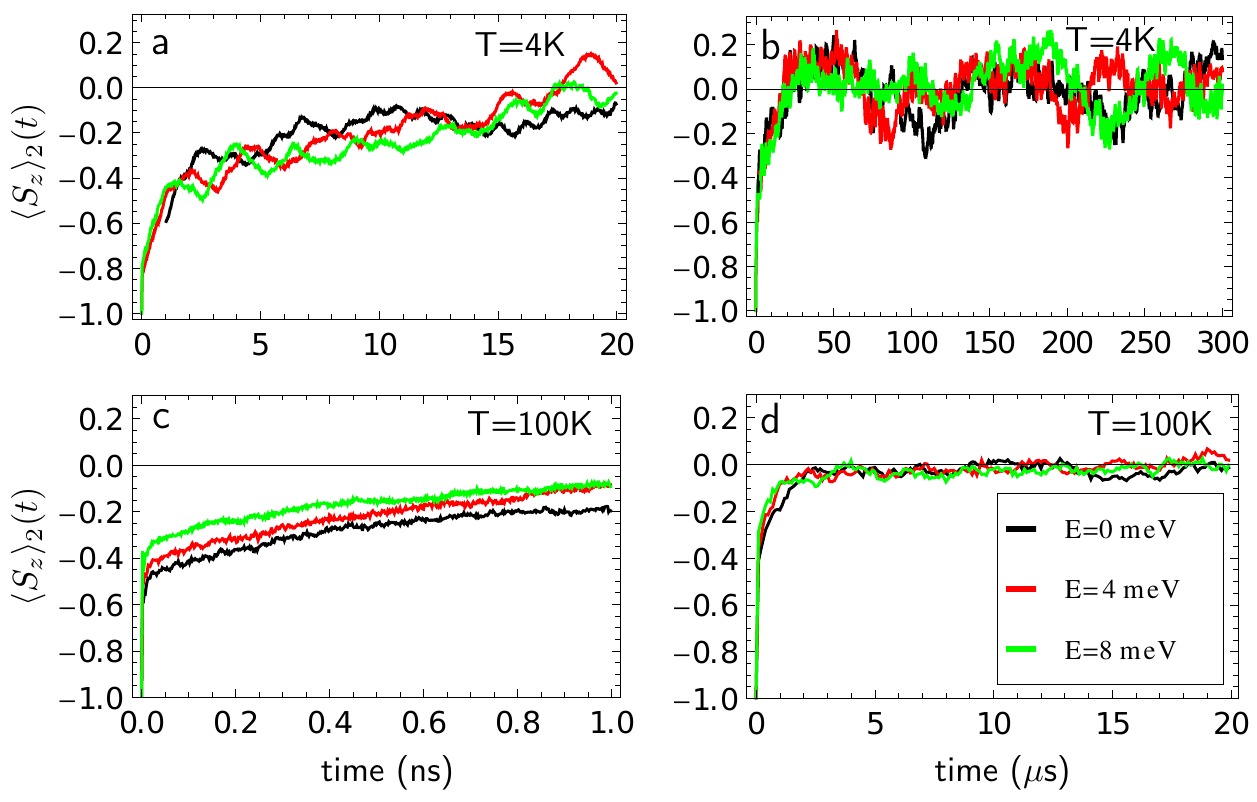}
\caption{Return to equilibrium for $S_z$ of a single adatom spin coupled to 8 substrate electrons. The relaxation is shown for temperatures of $T=4$ K, $T=100$ K,
different values of $E$ and fixed $D=1$ meV. Fig. 3 a) and 3 c):
The short time behavior shows a faster relaxation for a higher temperature. Fig. 3 b) and 3 d): The long time behavior shows smaller quantum
fluctuations around thermal equilibrium for higher temperatures. The different time scales on the $x$-axis should be noted.}
\end{figure}

\begin{figure}[h]
 \includegraphics*[scale=0.68,viewport=149 441 511 672]{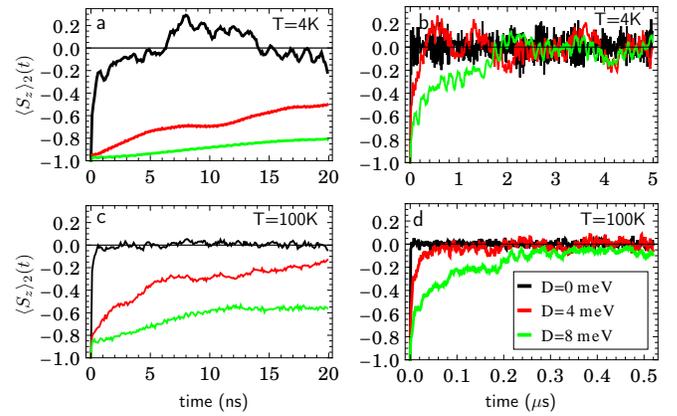}
\caption{Return to equilibrium for $S_z$ and different values of $D$ of a single adatom spin coupled to 8 substrate electrons.
The relaxation time grows with increasing value of the anisotropy barrier $D$.
Fig. 4a) and 4c): The short time behavior shows a faster
relaxation for a higher temperature. Fig. 4b) and 4d): The long time behavior shows smaller quantum fluctuations around thermal equilibrium
for higher temperatures. The different time scales on the $x$-axis should be noted.}
\end{figure}
It can be seen that in all cases the fluctuations decrease with increasing temperature. Depending on the time scale, in all cases the evaluated function
$\omega^{\beta P}(\tau_t(S_z))$ can be approximated by a function starting from $-1$ with an exponential decay to zero. Fluctuations induced by the temperature are not
able to switch the spin back to $\langle S_z\rangle\approx-1$. A similar behavior has been found experimentally in \cite{LothIBM}, where a single Fe spin was excited
with a high voltage pump, corresponding to a strong perturbation in our model, and the relaxation process of this single spin was investigated.
The magnetization showed the exponential decay for first time period, followed by small fluctuations near thermal equilibrium. The temperature was unable
to switch the spin to its initial value for $t=0$.
The appearance of larger fluctuations for a lower temperature
might be explained by energy considerations. A system at a lower temperature has less energy than a system at a higher temperature.
A perturbation $P$ corresponds to an additional amount of energy.
Notice that for an infinite system this additional energy is negligible and an exact thermalization might take place \cite{Bratelli}.
The relative ratio between the energy of perturbation and the energy of the free system is larger at lower temperatures.
This might be a reason for the stronger fluctuations at lower temperatures.
Another important effect at low temperatures are "quantum fluctuations", which become extinct with increasing temperature.
We can also see that for higher temperatures the quantum
spin of the adatom returns faster to equilibrium, i.e. the adatom relaxation time becomes shorter. With increasing value of $D$ the relaxation time also increases.
This is in agreement with the statement that a spin "up" or "down" state becomes more stable with increasing anisotropy barrier.
For increasing $E$ (see Fig. 3 a,c) an inverted behavior has been observed for short times.

Now we will apply Eq. (\ref{eq:measure}) and (\ref{eq:measurement}) to model the STM measurement process as a time average.
As an example we take cobalt adatoms on platinum (111) \cite{Meier:Science(2008)}.
As described in the introduction the reason for zero magnetization of a Co adatom on Pt(111) at zero external field was still unclear (see Fig. 5).
The Hamiltonian for the cobalt atom is given by \cite{Meier:Science(2008)}
\begin{equation}
 H=-m B_z S_z-K S^2_z,
\end{equation} with $m=3.7$ $\mu_B$ and $K=9$ meV.
From this Hamiltonian a nearly vanishing probability for the states $|-\frac{3}{2}\rangle$, $|-\frac{1}{2}\rangle$, $|\frac{1}{2}\rangle$ and $|\frac{3}{2}\rangle$ can be found in thermal equilibrium.
\begin{figure}[h]
 \includegraphics*[scale=0.75,viewport=125 533 451 672]{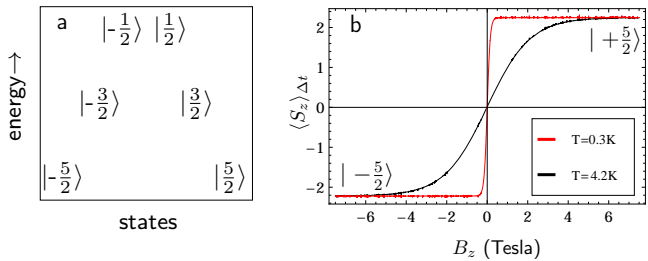}
\caption{a) The high energy states $|-\frac{3}{2}\rangle$, $|-\frac{1}{2}\rangle$, $\mid\frac{1}{2}\rangle$ and
$\mid\frac{3}{2}\rangle$ got a vanishing probability in thermal equilibrium. The preferred states are those where the spin points "up" $\mid\frac{5}{2}\rangle$
or "down" $\mid-\frac{5}{2}\rangle$. b) The time averaged magnetization curves for two different temperatures show agreement of Eq. (\ref{eq:measurement})
with the experimental data in \cite{Meier:Science(2008)}. For high positive magnetic field mostly the
$\mid\frac{5}{2}\rangle$ state is occupied and for high negative field it is the $\mid-\frac{5}{2}\rangle$ state.}
\end{figure}

If the spin has been prepared to be polarized in positive or negative $z$-direction it is not a priori clear how the spin can switch in the opposite
state because of the high anisotropy barrier.
The temperature of $T=0.3$ K and $T=4.2$ K used in this experiment is much too low to switch the spin over the anisotropy barrier of $K=9$ meV. The
N\'{e}el-Brown law predicts a switching time of a few million years, which is in disagreement with the short resolution time of $10$ ms of the SP-STM technique used in
\cite{Meier:Science(2008)}. The absence of the transverse anisotropy term $E(S^2_x-S^2_y)$ in the free system prevents direct transitions under the barrier between the
$|\frac{5}{2}\rangle$ and $|-\frac{5}{2}\rangle$ states.
To explain the zero expectation value of $S_y$ a quantum tunneling or a current induced magnetization switching mechanism has been speculatively proposed \cite{Meier:Science(2008)}.
Here, we check this proposition by numerical calculations.

The perturbation is taken to be $P=J\sum_i\vec{S}\vec{\sigma}_i+\sum_i\vec{m}_{\mathrm{tip}}\vec{\sigma}_i$,
with the magnetization $\vec{m}_{\mathrm{tip}}$ of the tip and the Pauli matrices corresponding to the tunneling electrons.
When the cobalt atom gets perturbed because of the interaction with the tunneling electrons it gets out of equilibrium and
the question about the occupation probability of the states $|\frac{5}{2}\rangle$,...,$|-\frac{5}{2}\rangle$ arises.
A related question is, in which way the spin gets from the
$|\frac{5}{2}\rangle$ to the $|-\frac{5}{2}\rangle$ state. Especially interesting is the case of zero magnetic field where the SP-STM measurement provides
a time averaged expectation value $\langle S_z\rangle_{\Delta T}=0$.
Fig. 6 a), b) gives the time evolution $\omega^\beta(\tau^P_t(S_z))$ for the z-component of the magnetization, for two different values of external magnetic field $B_z$.
In agreement with experimental data $\langle S_z\rangle_{\Delta t}=0$ for $B_z=0$, while it increases with increasing $B_z$.
\begin{figure}[h]
 \includegraphics*[scale=0.6,viewport=149 413 554 672]{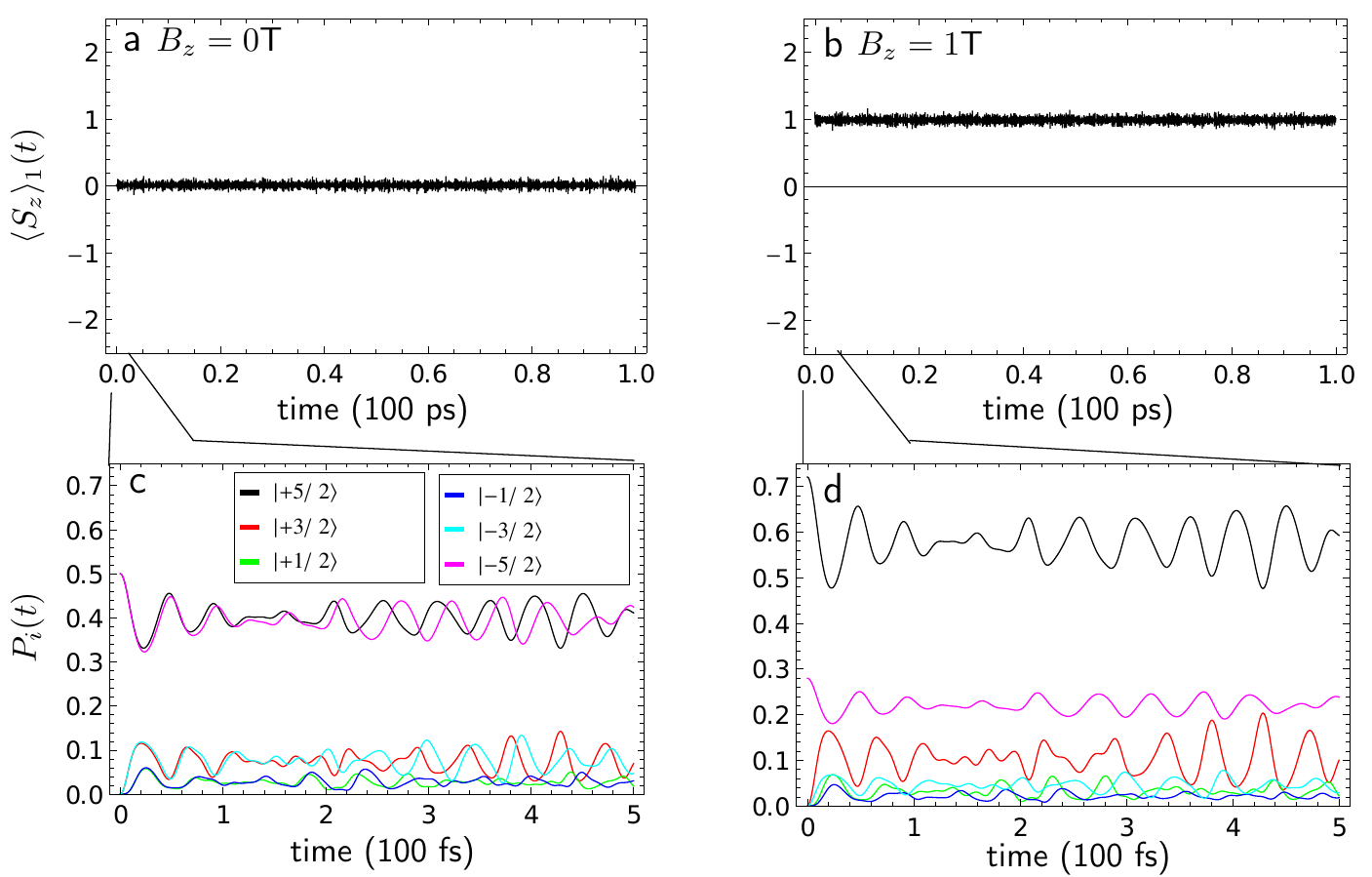}
\caption{The time evolution of the adatom spin for the $z$-component as a function of time are shown for a), c) $B_z=0$ and
b), d) $B_z=1$ Tesla.
The expectation value fluctuates around thermal equilibrium at $T=4.2$ K.
The occupation probabilities $P_i$ for the states $i=|\frac{5}{2}\rangle$,...,$|-\frac{5}{2}\rangle$ as a function of time are
shown for c) $B_z=0$ and d) $B_z=1$ Tesla.}
\end{figure}
As it is seen in Fig.6 c) at $t=0$ the total signal is composed of the superposition of $|\frac{5}{2}\rangle$ and $|-\frac{5}{2}\rangle$.
As the tunneling current is switched on the occupation probabilities of those states start to oscillate. The amplitude of oscillations
increases with increasing parameter $J$ and also depends on $\vec{m}_{\mathrm{tip}}$.
The appearance of fluctuations in the occupation probabilities means that not only $\mid\pm\frac{5}{2}\rangle$ states become occupied. In other words magnetization
switching occurs. The results demonstrate that
even a weak perturbation due to tunneling electrons initiates a quantum tunneling in otherwise diagonal systems.
The expectation value of magnetization at $B_z=0$ remains nearly zero.

\section{Conclusion}
The non-equilibrium dynamics of adatoms on different substrates under the action of a magnetic STM tip has been studied.
A satisfactory agreement of Eq. (\ref{eq:r}) and (\ref{eq:measure}) with experimental data has been found, when the perturbation has been identified as the interaction
between the STM-tip and the sample. It has been shown that the application of a perturbed KMS state in (\ref{eq:r}) is well suited to model relaxation dynamics of
magnetic atoms at finite temperature.
The application of the perturbed dynamics in Eq. (\ref{eq:measure}) can be used to investigate the system dynamics during an SP-STM measurement.

Fig. 2, 3 and 4 demonstrate that thermalization can be achieved for relatively small systems, which can be calculated using exact diagonalization.
The relaxation can be approximated with an exponential function, which is in agreement with experimental results.
It is demonstrated that the lifetime of single adatoms increases with increasing anisotropy barrier and decreasing temperature.

We were able to reproduce the experimentally obtained time averages of expectation values.
Additionally the dynamics of the sample (see Fig. 6) for time scales which are
shorter than the resolution time of the STM has been described using
the integrand of Eq. (\ref{eq:measurement}).

A finer mathematical structure \cite{Bratelli} can be implemented into the rough structure described in section II.
This is done by the replacement of the substitution of the Hamiltonians $H\rightleftarrows H+P$ by $C^*$-dynamical systems $(\mathcal{A},\tau)\rightleftarrows(\mathcal{A},\tau^P)$.
This general mathematical structure can also be applied to fermionic lattice systems, e.g. the Hubbard model, and continuous fermionic systems in the algebraic
approach to Quantum Field Theory.

\acknowledgments
Support by the DFG (SFB 668, project A11), by the Hamburg Cluster of Excellence "Nanospintronics" and by the ERC Advanced Grant "FURORE" is gratefully acknowledged.

\end{document}